# g-SiC$_6$ Monolayer: A New Graphene-like Dirac Cone Material with a High Fermi Velocity

Tao Yang [a,b], Xingang Jiang [a], Wencai Yi [*a], Xiaomin Cheng [*b] and Xiaobing Liu[*a]

a.  *Laboratory of High Pressure Physics and Material Science (HPPMS), School of Physics and Physical Engineering, Qufu Normal University, Qufu, Shandong, 273165, China*

b.  *Institute of Advanced Materials, School of Electromechanical and Automobile Engineering, Huanggang Normal University, Huanggang, Hubei, 438000, China*

**ABSTRACT**

Two-dimensional (2D) materials with Dirac cones have been intrigued by many unique properties, i.e., the effective masses of carriers close to zero and Fermi velocity of ultrahigh, which yields a great possibility in high-performance electronic devices. In this work, using first-principles calculations, we have predicted a new Dirac cone material of silicon carbide with the new stoichiometries, named g-SiC$_6$ monolayer, which is composed of *sp$^2$* hybridized with a graphene-like structure. The detailed calculations have revealed that g-SiC$_6$ has outstanding dynamical, thermal, and mechanical stabilities, and the mechanical and electronic properties are still isotropic. Of great interest is that the Fermi velocity of g-SiC$_6$ monolayer is the highest in silicon carbide Dirac materials until now. The Dirac cone of the g-SiC$_6$ is controllable by an in-plane uniaxial strain and shear strain, which is promised to realize a direct application in electronics and optoelectronics. Moreover, we found that new stoichiometries AB$_6$ (A, B = C, Si, and Ge) compounds with the similar SiC$_6$ monolayer structure are both dynamics stable and possess Dirac cones, and their Fermi velocity was also calculated in this paper. Given the outstanding properties of those new types of silicon carbide monolayer, which is a promising 2D material for further exploring the potential applications.

# 1. INTRODUCTION

Dirac materials possess many significant properties, such as ultrahigh carrier mobility and ballistic charge transport,[1-3] which mostly due to the linear energy dispersion in the Fermi level.[4] The typical Dirac materials is graphene, the Fermi velocity is up to $8.3 \times 10^5$ m/s,[5] and is considered to be a promising candidate for high-performance electronic devices in future technologies.[6,7] Recently, many new two-dimensional (2D) Dirac materials have been identified and have received wide attention, such as carbon allotropes,[8,9] silicene,[10] germanene,[11] $SiC_3$,[12] $B_2Se$,[13] BP,[14] and honeycomb boron.[15] Among them, the compounds composed with Group IV elements have the most important structural prototypes as Dirac materials, the graphene,[5,7,16] silicene,[10,17] germanene,[11,18] and stanene[19] are Dirac cone materials with a strictly single element and hexagonal configurations. Based on the structural features of these Dirac materials, many rules have been proposed to explain the origin of Dirac cones and predict new Dirac materials,[20,21] i.e., Liu et al. have proposed "triple coupling" and "ring coupling" mechanism to elucidate the Dirac cones formation mechanism.[20]

By substituting carbon atoms of graphene with silicon atoms, many new silicon carbides were predicted and show intriguing electronic properties. However, most of them are semiconductors with bandgap, like g-SiC,[22] $g-SiC_2$,[23] and $g-SiC_7$[24] only the $g-SiC_3$ monolayer is a Dirac cone material.[12,20] A question naturally arises: can other stoichiometries of silicon carbide form new Dirac cone materials? We noticed that a recent study has reported that the $PC_6$ monolayer was composed of $C_6$ ring basic blocks and present a direct-gap semiconductor with a bandgap of 0.84 eV.[25] The presence of lone electron pairs in P atoms restrains the formation of the extended $\pi$ electron states of the ethylene-like C-C bond, which is in sharp contrast to graphene. Thus, could we obtain a new $SiC_6$ monolayer with conjugated $\pi$ electron states by substituting P atoms with silicon atoms in $PC_6$ monolayer?

In this work, we report a new 2D Dirac cone material of silicon carbide of $g-SiC_6$ monolayer, which is composed of *sp*$^2$ hybridized with a graphene-like hexagonal honeycomb lattice configurations. The detail calculations reveal that $g-SiC_6$ monolayer is with outstanding dynamical and thermal stabilities, and possesses a Dirac cone at the high-symmetry *K* point with a high Fermi velocity. This Dirac cone could be controlled by the in-plane uniaxial strain and shear strain. More interesting, the new stoichiometries $AB_6$ (A, B = C, Si, and Ge) compounds are predicted and they all belong to new Dirac materials by substituting the elements of $g-SiC_6$ structural prototype.

## 2. COMPUTATIONAL DETAILS

All calculations were carried out using the density functional theory, implemented in *Vienna Ab-Initio Simulation Package* (VASP).[26-28] The exchange-correlation potential was employed by the generalized gradient approximation (GGA) with Perdew-Burke-Ernzerh (PBE) functional.[29,30] The cutoff energy and the convergence of total energy were set to be 700 eV and $10^{-5}$ eV, respectively. To eliminate the interactions between neighboring layers, the 24 Å vacuum layer thickness was used along the *c* axis in our simulations. Uniform G-centered k-grid mesh with a resolution of $11 \times 11 \times 1$ in the Brillouin zone for structure optimization and a resolution of $32 \times 32 \times 1$ k-grid mesh was used in electronic properties calculations. The structure relaxation proceeded until the residual Hellmann-Feynman forces on atoms were less than 0.01 eV/Å. To explore the dynamic stability, the vibrational properties were investigated using the PHONOPY code[31] with the forces calculated from VASP. The Heyd-Scuseria-Ernzerhof (HSE06) hybrid functional was employed to calculate the electronic properties, which was used to confirm the Dirac cones.[32] In the molecular dynamics (MD) simulations, the initial configuration in the supercell is annealed at different temperatures, each MD simulation in the NVT ensemble lasts for 6 ps with a time step of 2.0 fs, and the temperature is controlled by using the Nosé-Hoover method. The post-processing of VASP calculated data used qvasp code.[33]

## 3. RESULTS AND DISCUSSION

After geometric optimization, unlike $PC_6$ monolayer, the structure of g-$SiC_6$ monolayer is perfectly planar without any buckling in Figure 1a, and the calculated structural parameters of relative compounds are summarized in Table 1. This g-$SiC_6$ monolayer is 0.33 eV per unit lower than previously reported non-planar semiconducting 2D $SiC_6$.[34] Figure 1b shows the electron localization function (ELF) of the $SiC_6$ monolayer. The ELF values between nearest-neighbor C atoms and Si and C atoms are close to 0.82 and 0.85, respectively, indicate that there formed strong covalent bonds in g-$SiC_6$ monolayer. The primitive cell of g-$SiC_6$ is composed of the basic building block of $SiC_3$ and $C_6$ unit (Figure 1c). The bond length of the C-C bond is 1.455-1.465 Å, which is a bit longer than $sp^2$ C-C bond length in graphene (1.42 Å). The Si-C bond length of 1.731 Å is comparable to g-$SiC_2$ (1.789 Å) and g-$SiC_7$ (1.69 Å). Then, we analyze the bonding characteristics and bonding strength of C-C and Si-C bonds in the g-$SiC_6$ monolayer by the Crystalline Orbital Hamiltonian Population (COHP). For comparison, the -ICOHP of chemical bonds of PC$_6$ monolayer was calculated. As shown in Figure 1d, the results show that most valance electrons occupied bonding states, and only a slight occupancy at antibonding states near the Fermi level in C-C bond. Moreover, the calculated results of -ICOHP revealed that all C-C bonds in the g-SiC$_6$ monolayer are beyond 9.256, which means that there formed a very stable covalent bond between C atoms. The -ICOHP of all Si-C bonds are beyond 6.581, which is a bit larger than the P-C bond in the PC$_6$ monolayer (6.430). Therefore, those results indicated that g-SiC$_6$ is a relatively stable system. Overall, these bonding configurations inevitably enhance the structural stability of g-SiC$_6$ monolayer.

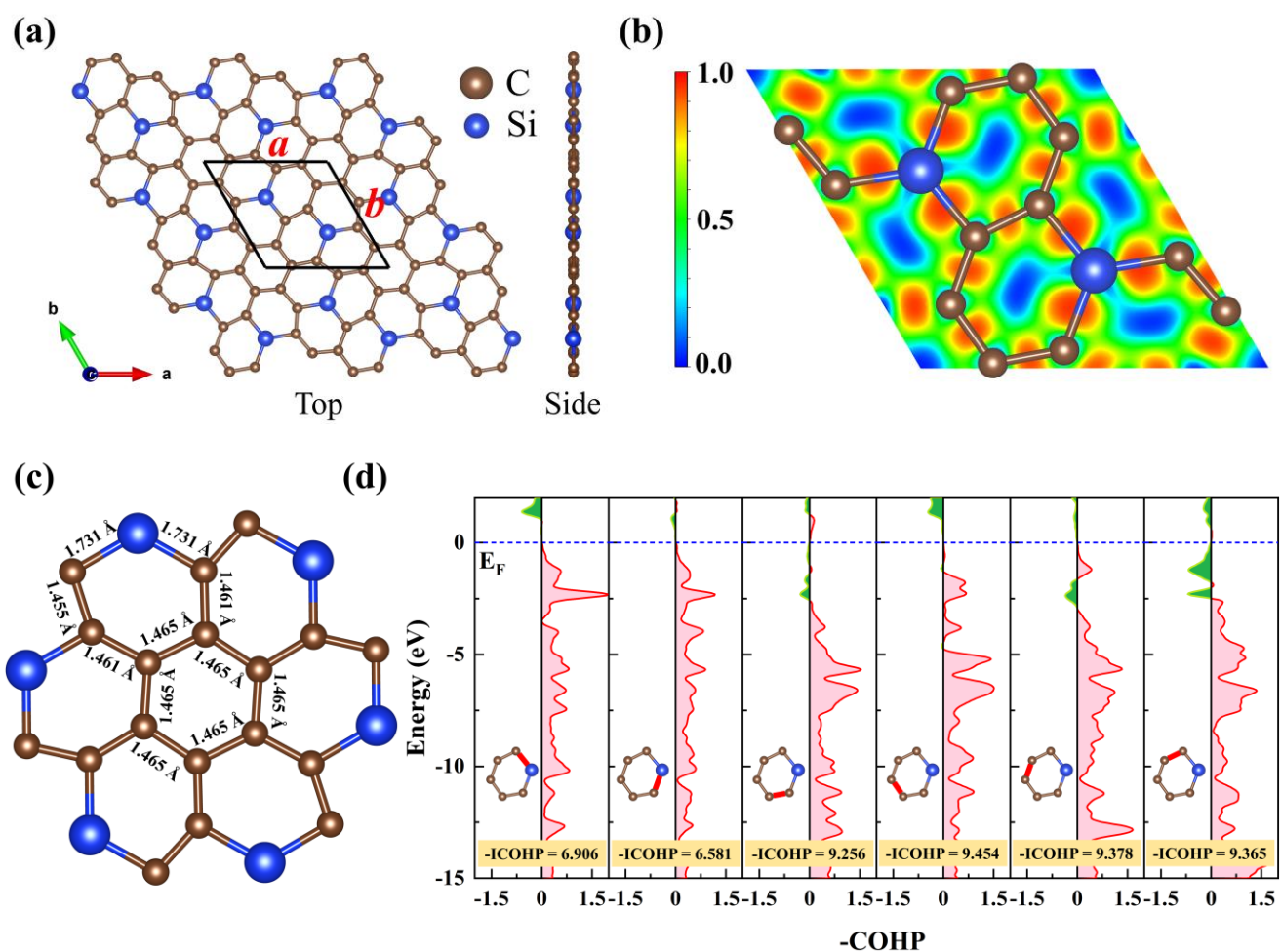

**Figure 1.** (a) Optimized geometrical structure of g-SiC$_6$ monolayer was viewed from different directions. (b) ELF maps of the g-SiC$_6$ monolayer. Blue and brown spheres denote Si and C atoms, respectively. (c) The bond length information of g-SiC$_6$ monolayer. (d) The -COHP and -ICOHP value for chemical bonds (marked as red) in the g-SiC$_6$ monolayer. The pink corresponds to bonding states, and green corresponds to antibonding states.

**Table 1.** Calculated properties of the g-SiC$_6$ monolayer, including the Space Group, Lattice Constants (Å), Bond Lengths (Å), Band Gaps $E_g$ (eV), and Fermi velocity (10$^5$ m/s). The relative materials are also listed for comparison. [35]

| Model | Space Group | Lattice Constants (Å) | Bond Length (Å) C-C | Bond Length (Å) Si-C | $E_g$ (eV) | Fermi velocity (10$^5$ m/s) |
|---|---|---|---|---|---|---|
| Graphene | $P\bar{3}m1$ | 2.468 | 1.424 | - | 0 | 8.2 (PBE)[5] <br> 10.5 (HSE06) |
| g-SiC | | 3.094 | - | 1.786 | 2.53 (PBE) [22] | - |
| g-SiC$_2$ | $P\bar{6}2m$ | 5.019 | 1.445 | 1.789 | 0.6 (PBE) <br> 1.09 (HSE06) [23] | - |
| g-SiC$_3$ | $P6/mmm$ | 5.633 | 1.438 | 1.813 | 0[12] | 6.0 (PBE)[20] |
| g-SiC$_6$ | $P\bar{3}$ | 7.008 | 1.455-1.465 | 1.731 | 0 | 7.11 (PBE) <br> 9.31 (HSE06) |
| g-SiC$_7$ | $P\bar{6}m2$ | 5.273 | 1.436-1.550 | 1.691 | 0.76 (PBE) <br> 1.13 (HSE06) [24] | - |

To evaluate the stability of g-SiC$_6$ monolayer, the phonon spectrum was calculated and was shown in Figure 2a. Obviously, there are no imaginary frequency of lattice vibrations in the whole Brillouin zone, suggesting that the g-SiC$_6$ monolayer has good dynamic stability. Further examination of the thermal stability is performed by using ab initio molecular dynamics (MD) simulations. The 3 × 3 supercell is adopted by heating the structure to 300 K and 600 K within a canonical ensemble. At the end of simulation, the final structure is carefully examined. The evolution of total energy as a function of time are presented in Figure 2b. At 300 K, the total energy of g-SiC$_6$ monolayer could fluctuate around -1033.85 eV, and the structural skeleton could be maintained well. Even the temperature increases to 600 K (see Figure S1), it can still withstand slight distortions that are not sufficient to destroy the Si-C and C-C bonds. Thus, the results of the phonon spectrum and MD simulations indicate the highly dynamic and thermal stability of g-SiC$_6$ monolayer.

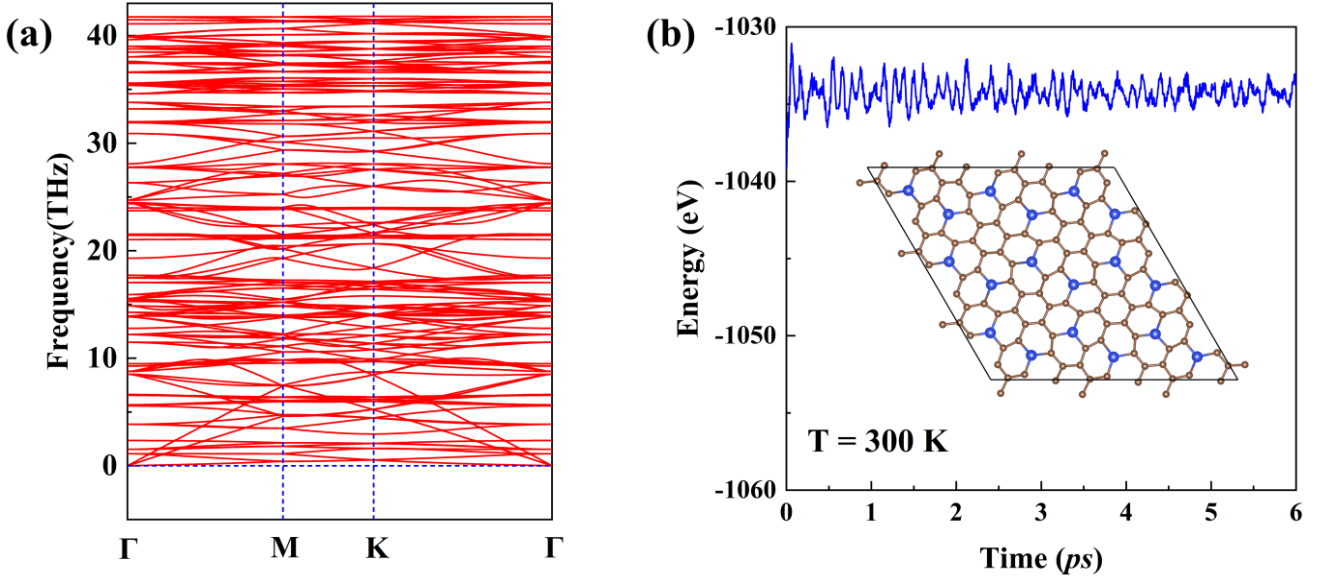

**Figure 2.** (a) Phonon spectrum of the g-SiC$_6$ monolayer. (b) Snapshots of g-SiC$_6$ monolayer at the temperatures of 300 K after 6 ps MD simulations.

Here, we examined the mechanical stability of SiC$_6$ by calculating the linear elastic constants. The calculated results of 2D linear elastic constants are as follows: $C_{11}$ = 275.873 N/m, $C_{22}$ = 275.912 N/m, $C_{12}$ = 69.141 N/m, $C_{66}$ = 103.448 N/m. For an oblique lattice of 2D materials, the stability standard are as follows: $C_{11} > 0$, $C_{11}C_{22} > C_{12}^2$, det $(C_{ij}) > 0$.[36] Therefore, all the results satisfy the conditions mentioned above, meaning that the g-SiC$_6$ monolayer is mechanically stable. Besides, we further evaluated Young's modulus $Y(\theta)$ and Poisson ratio $v(\theta)$ along an arbitrary direction $\theta$ ($\theta$ is the angle relative to the positive $x$-direction of the primitive cell) can be expressed as:

$$Y(\theta) = \frac{C_{11}C_{22} - C_{12}^2}{C_{11}s^4 + C_{22}c^4 + \left(\frac{C_{11}C_{22} - C_{12}^2}{C_{44}} - 2C_{12}\right)c^2s^2}$$

$$v(\theta) = \frac{\left(C_{11} + C_{22} - \frac{C_{11}C_{22} - C_{12}^2}{C_{44}}\right)c^2s^2 - C_{12}(c^4 + s^4)}{C_{11}s^4 + C_{22}c^4 + \left(\frac{C_{11}C_{22} - C_{12}^2}{C_{44}} - 2C_{12}\right)c^2s^2}$$

where c = cos($\theta$) and s = sin($\theta$). The calculation results display that the Young's modulus and the Poisson ratio are both isotropic, and the values are 258.547 N/m and 0.251, respectively. Comparing with other typical 2D materials, the in-plane Young's modulus of g-SiC$_6$ monolayer is close to BN

(~267 N/m),[37] is smaller than graphene (~340 ± 40 N/m)[38] and BC$_3$ (~316 N/m).[39] However, the g-SiC$_6$ monolayer is still much harder than the experimentally synthesized silicene (~62 N/m).[40] These results implying that the g-SiC$_6$ monolayer has strong mechanical strength among 2D materials.

We further focused on the electronic property of g-SiC$_6$ monolayer. As shown in Figure 3a and 3b, the band structure was calculated by both the PBE functional and the HSE06 functional and the results demonstrate that the g-SiC$_6$ monolayer is a new Dirac material. The valence and conduct bands intersect linearly at the Fermi level, forming a zero-gap Dirac cone at the *K* point. In view of the PDOS (Figure 3a and Figure S2), we can find that the bands around the Dirac cone are occupied by only the $p_z$ orbitals from both the Si and C atoms, forming $\pi$ and $\pi^*$ bonds, which is similar to graphene.

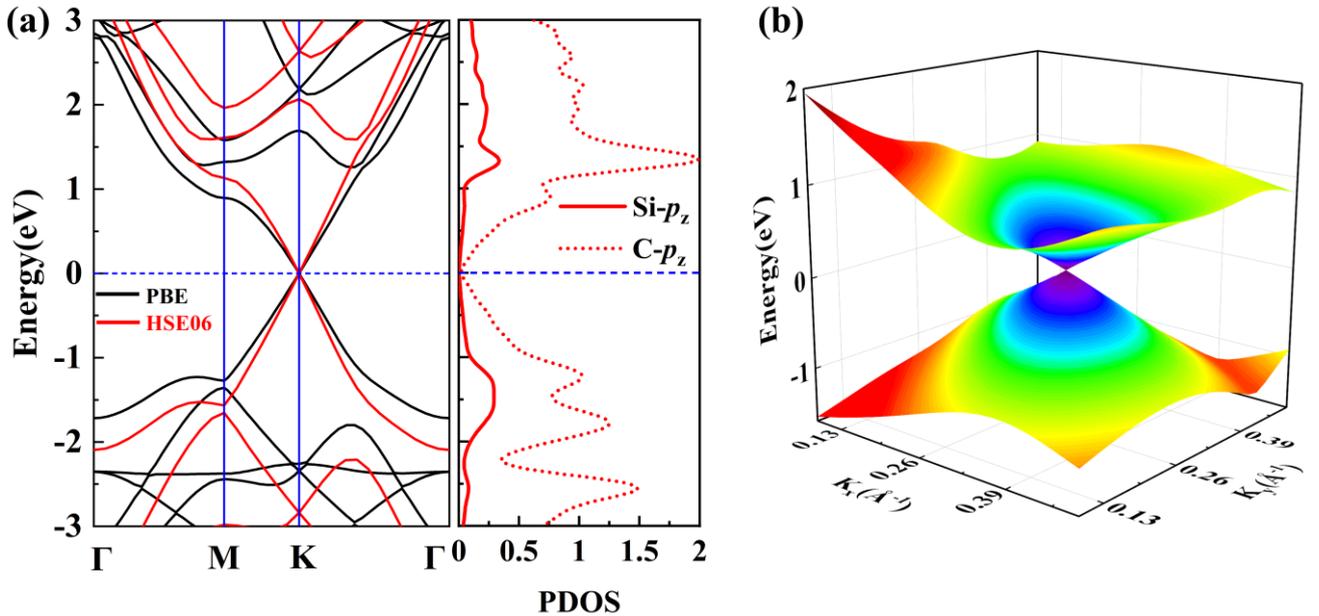

**Figure 3.** (a) The band structure of the g-SiC$_6$ monolayer, was calculated by both PBE (black lines) and HSE06 (red lines) functional. The Fermi level is set to zero and denoted by blue dashed lines. (b) Dirac cone formed by the valence and conduction bands in the vicinity of the Dirac point of the g-SiC$_6$ monolayer.

The Fermi velocity is one of the most important features for Dirac materials, the linear dispersion of energy bands suggests the zero-effective mass of the carriers near the Fermi level. As shown in Figure S3, the arrangements of Si and C atoms in the directions are clearly different. Thus, we use a large supercell with 196 atoms to calculate the band structures along the armchair ($k_x$) and zigzag ($k_y$) directions. As shown in Figure S4, the valence and the conduction bands in the vicinity of the Dirac point suggest the presence of a Dirac cone in the large supercell. The slopes of the bands along the $k_x$ and $k_y$ directions are ± 29.43 and ± 29.72 eV·Å, respectively, implying direction-independent Fermi velocities in the g-SiC$_6$ monolayer. By fitting the linear energy bands around the Dirac point, we could obtain the Fermi velocity of the Dirac Fermions according to the equation $v_f = \frac{1}{\hbar}\frac{\partial E}{\partial k}$. It is found that the Fermi velocity is predicted to be 7.11 × 10$^5$ m/s by using the PBE functional, which close to that of graphene (8.2 × 10$^5$ m/s) [5], and is larger than that of many other Dirac materials, i.e., g-SiC$_3$(6.0 × 10$^5$ m/s). The high Fermi velocity is greatly beneficial for high-performance electronic devices.

For examining the symmetrical protection of the Dirac cone, in-plane biaxial and uniaxial strains are applied to investigate the effects on the band structure of g-SiC$_6$ monolayer. The corresponding results are shown in Figure 4. Obviously, the Dirac cone can still be preserved under the biaxial strain (Figure 4a). A small difference is that the Fermi velocity of the g-SiC$_6$ monolayer is strongly dependent on the in-plane biaxial strains. For instance, the biaxial strain can slightly change the Fermi velocity from 7.11 × 10$^5$ m/s to 6.51 × 10$^5$ m/s (5%) and 7.61 × 10$^5$ m/s (-5%). All of these velocities are on the same order of magnitude and comparable of that of graphene, suggesting that the in-plane biaxial strains have no significant influence on Fermi velocity. For the uniaxial strain (Figure 4b and c), a magnitude of ±5% strain can open a little gap (22.0 meV~37.5 meV) along the armchair or zigzag direction. This is very similar to the graphene cases where a uniaxial strain of more than 26.2% can open a bandgap of 45.5 meV.[41] However, an in-plane shear strain can open a bandgap of 0.22 eV easily when a magnitude of 5% strain is applied (see Figure S5 and S6), which is much smaller than that of graphene (~16%).[42] All our calculations demonstrated that the symmetry protection of the Dirac cone in g-SiC$_6$ monolayer could be easily controlled and the bandgap could be opened by strain, which makes g-SiC$_6$ monolayer have a potential application in electronics and optoelectronics.

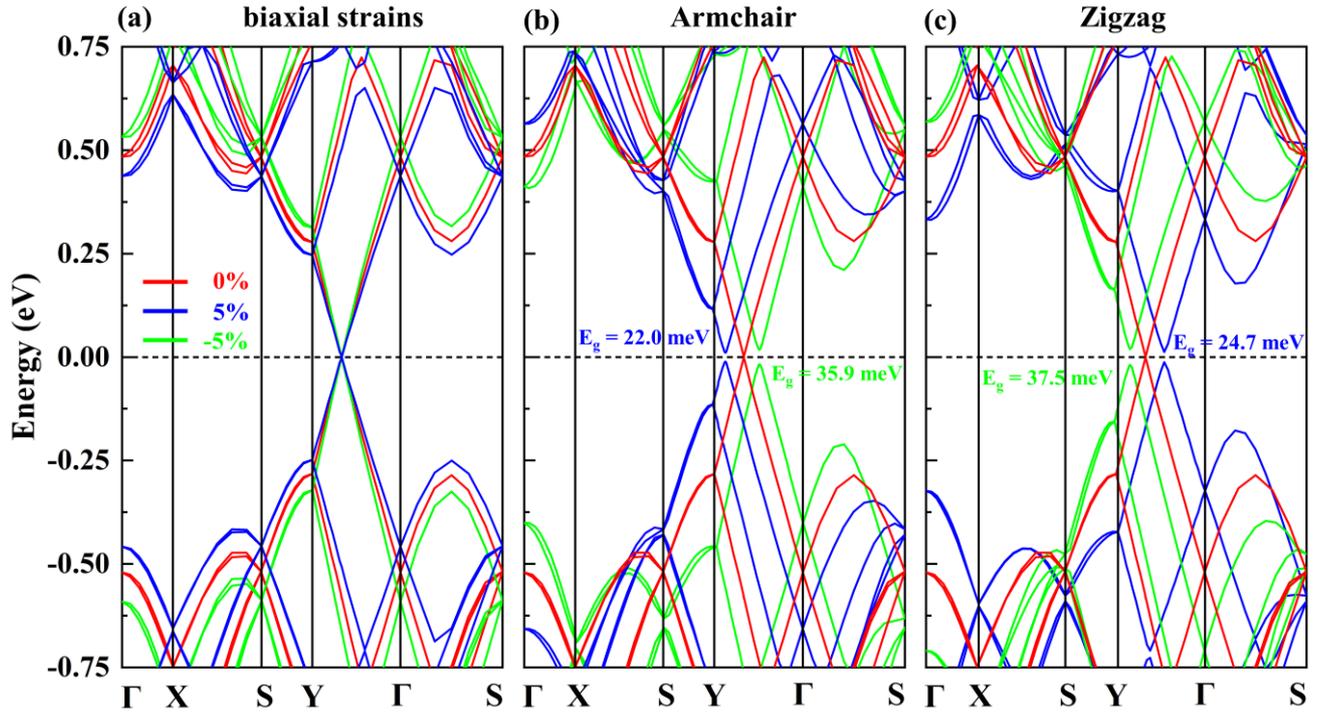

**Figure 4.** Variations of electronic band structure against different in-plane strains: (a) biaxial and uniaxial strains along (b) armchair (*y* axis) and (c) zigzag (*x* axis) directions. Green lines: -5%. Red lines: 0%. Blue lines: 5%.

According to the "ring coupling" mechanism and isoelectronic rule, we further extend g-SiC$_6$ monolayer to other group-IV analogs g-AB$_6$ monolayer (A, B = C, Si, and Ge). Structural optimizations show that only the planar configuration is preserved in g-SiC$_6$ monolayer, whereas g-CSi$_6$, g-GeC$_6$, g-CGe$_6$, g-GeSi$_6$, and g-SiGe$_6$ monolayer have buckled structures with a thickness of 0.621-2.535 Å (Figure 5a). The buckling features of g-CSi$_6$, g-GeC$_6$, g-CGe$_6$, g-GeSi$_6$, and g-SiGe$_6$ are caused by the strong *sp*$^3$ hybridization preference of both Si and Ge atoms, which is similar with silicene and germanene. The structural parameters of these 2D frameworks are listed in Table 2. Although the center of inversion symmetry of the frameworks remains, even though the hexagonal symmetry is broken in buckled configurations. Thus, these g-AB$_6$ monolayers possess Dirac cone in their band structures (Figure 5b-g), and the Fermi velocity of AB$_6$ monolayer is the same magnitude order of that of graphene.

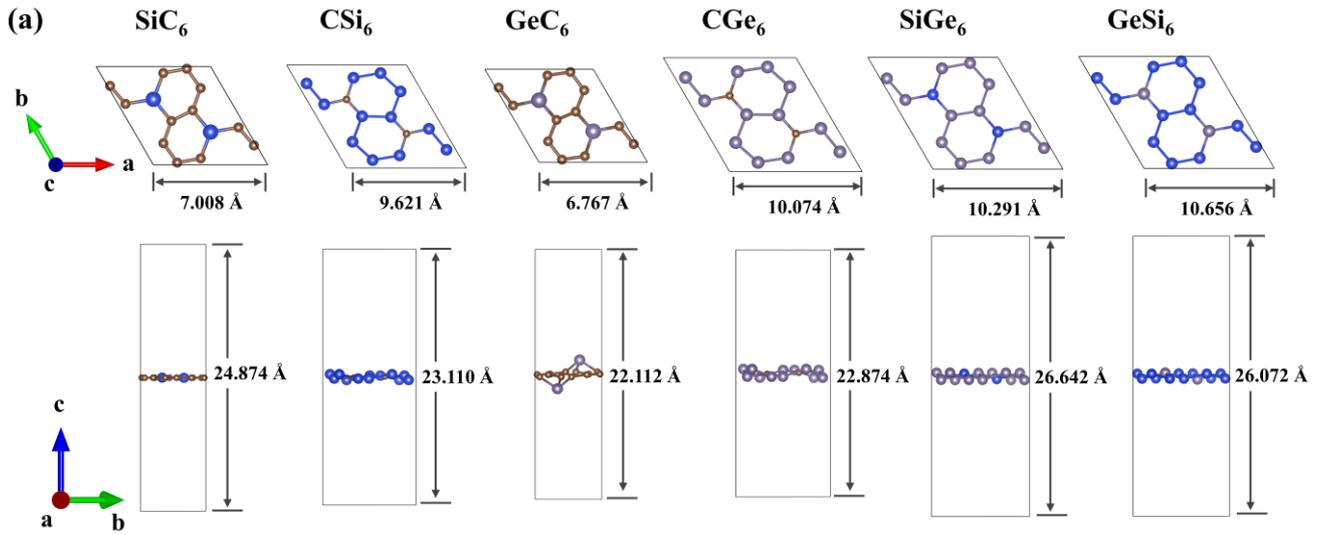
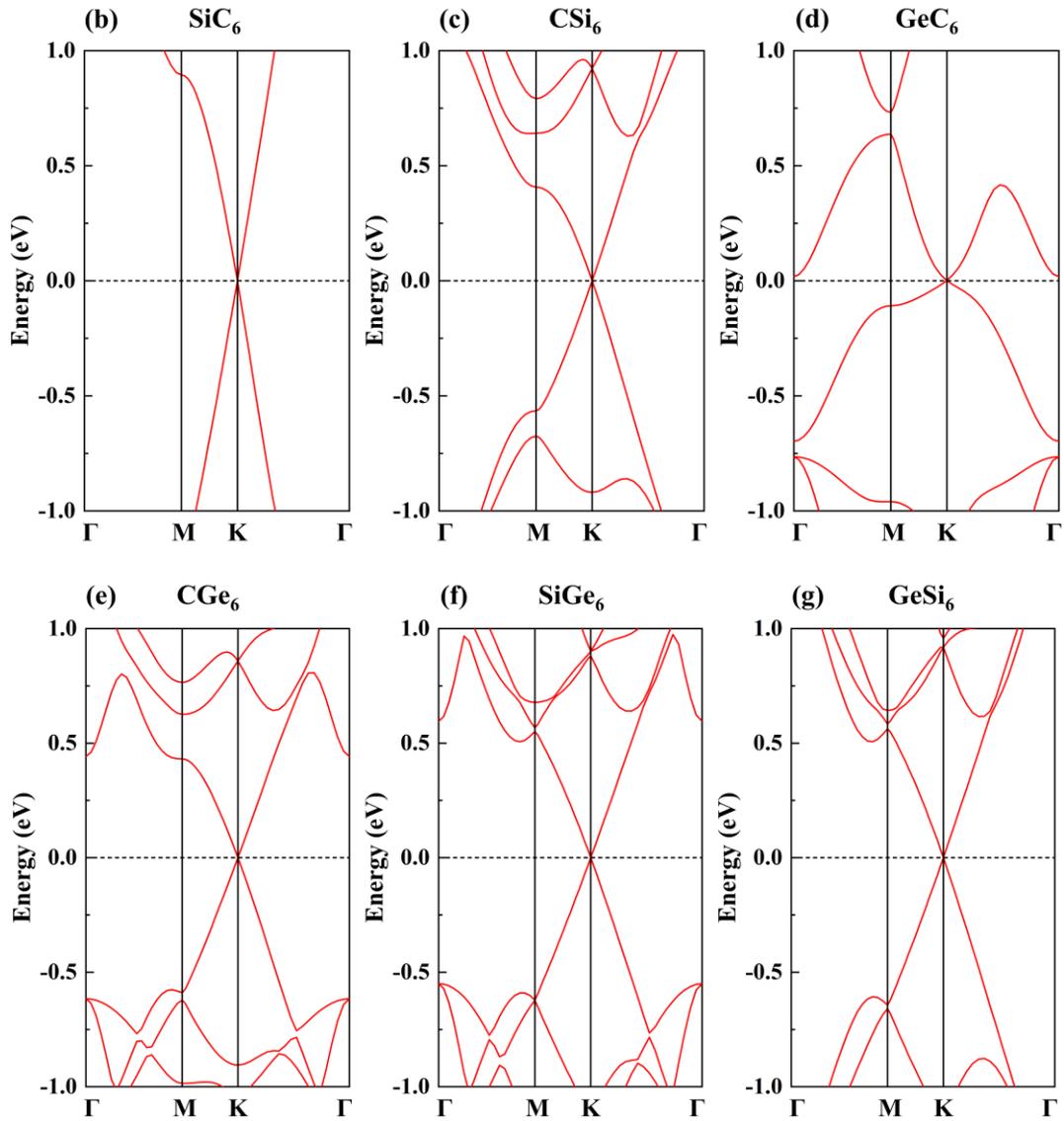

**Figure 5.** (a) The crystal structures of group IV analogs. The blue, brown, and purple colors correspond to Si atom, C atom, and Ge atom, respectively. Electronic band structure of the silicon

carbide analogs: (b) g-SiC6, (c) g-CSi6, (d) g-GeC6, (e) g-CGe6, (f) g-SiGe6, and (g) g-GeSi6. They are all Dirac cone materials.

**Table 2.** Calculated properties of group IV analogs $AB_6$ monolayer (A, B = C, Si, and Ge), including Lattice Constants (Å), Bond Lengths (Å), and Fermi velocity ($10^5$ m/s).

| Model | Lattice Constants (Å) | Bond Length (Å) | | Fermi velocity ($10^5$ m/s) |
| --- | --- | --- | --- | --- |
| | | C-C / Si-Si / Ge-Ge | Si-C / Ge-C / Ge-Si | |
| g-SiC6 | 7.008 | 1.455 ~ 1.465 | 1.731 | 7.11 |
| g-CSi6 | 9.621 | 2.235 ~ 2.340 | 1.843 | 5.30 |
| g-GeC6 | 6.767 | 1.374 ~ 1.463 | 1.955 | 1.53 |
| g-CGe6 | 10.074 | 2.405 ~ 2.528 | 1.953 | 4.98 |
| g-SiGe6 | 10.291 | 2.435 ~ 2.439 | 2.343 | 5.42 |
| g-GeSi6 | 10.656 | 2.280 ~ 2.284 | 2.374 | 5.35 |

To evaluate the relative stability of group IV analogs, we defined the cohesive energy $E_{co}$ as:

$$E_{co} = (E_{total} - N_A \times \mu_A - N_B \times \mu_B) / (N_A + N_B),$$

where $E_{total}$ is the total energy of $AB_6$ monolayer, and $N_A$ and $N_B$ are the numbers of group IV atoms in one supercell. The $\mu_A$ and $\mu_B$ are the chemical potentials of C, Si, and Ge atoms, for convenience, we reference the chemical potentials to the energy of graphene, cubic silicon crystal, and cubic Ge crystal, respectively. The $E_{co}$ of g-SiC6 monolayer is very close to graphene around group IV analogs (Figure 6), even better than previously reported of the g-SiC3.[20]

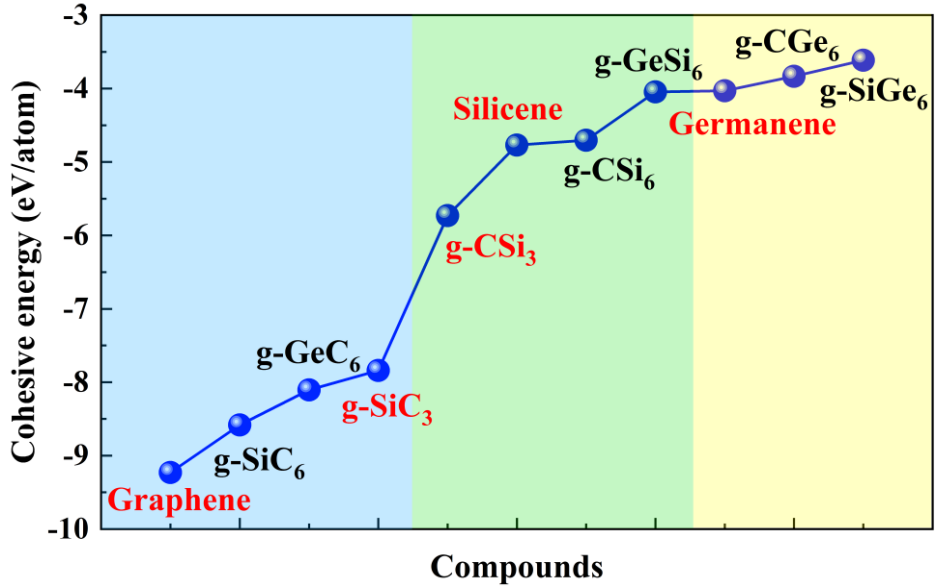

**Figure 6.** Formation energies of silicon carbide analogs. The values of graphene, g-SiC$_3$, g-CSi$_3$, silicene, and germanene are also proposed for comparison.[20]

We recognize that the realization of these metastable group IV analogs for experiments remains with a great challenge in the current technique. Fortunately, the recent progress on the growth of graphene on metal surfaces by chemical vapor deposition or organics precursor molecules may be helpful to achieve this goal. The controlled reaction between Si and graphene is another promising synthetic route to the siligraphenes as what happened in the growth of SiC nanotubes from carbon nanotubes.[43]

## 4. CONCLUSIONS

In conclusion, based on first-principles calculations, we have predicted a new 2D isotropic Dirac cone material of g-SiC$_6$ monolayer, which has been demonstrated to be dynamically, thermally, and mechanically stable. The Dirac cone in the g-SiC$_6$ monolayer is attributed to the $p_z$ orbitals of the Si and C atoms. The Fermi velocity ($\sim 10^5$ m/s) of g-SiC$_6$ monolayer is on the same order of magnitude as the graphene and also the highest in silicon carbide with the different stoichiometries. Interestingly, the bandgap could open under the in-plane uniaxial strain of 5% along the armchair or zigzag direction and in-plane shear strain of 5%, which is easier than the graphene (a uniaxial strain of ~26.2%, a shear strain of ~16%). In addition, the new structures group IV analogs of AB$_6$ monolayer still belong to Dirac materials, and g-SiC$_6$ monolayer has very low cohesive energy those new Dirac materials. All results for the Dirac cone and Fermi velocity render that ultrahigh carrier mobility can be achieved for

g-SiC$_6$ monolayer in the future electronics application.

**AUTHOR INFORMATION**

Corresponding Author

* E-mail: yiwc@qfnu.edu.cn (W. Y.)

* E-mail: chengxm@whut.edu.cn (X. C.)

* E-mail: xiaobing.phy@qfnu.edu.cn (X. L.)

**NOTES**

The authors declare no competing interest.

**ACKNOWLEDGEMENT**

This work was supported by the National Natural Science Foundation of China (Grants No. 21905159, 11974207 and 11974208), Natural Science Foundation of Shandong Province (Grants No. ZR2019BA010, ZR2019MA054, and 2019KJJ020) and the Project of Introduction and Cultivation for Young Innovative Talents in Colleges and Universities of Shandong Province. Calculations were performed at the High-Performance Computing Center (HPCC) of Qufu Normal University.

**REFERENCES**

[1] Castro Neto A H, Guinea F, Peres N M R, Novoselov K S, Geim A K. The electronic properties of graphene[J]. Reviews of Modern Physics, 2009, 81(1): 109-162.

[2] Bliokh Y P, Freilikher V, Nori F. Ballistic charge transport in graphene and light propagation in periodic dielectric structures with metamaterials: A comparative study[J]. Physical Review B, 2013, 87(24): 245134.

[3] Zhang Y, Tan Y-W, Stormer H L, Kim P. Experimental observation of the quantum Hall effect and Berry's phase in graphene[J]. Nature, 2005, 438(7065): 201-204.

[4] Geim A K, Novoselov K S: The rise of graphene, Nanoscience and Technology: Co-Published with Macmillan Publishers Ltd, UK, 2009: 11-19.

[5] Malko D, Neiss C, Viñes F, Görling A. Competition for Graphene: Graphynes with Direction-

Dependent Dirac Cones[J]. Physical Review Letters, 2012, 108(8): 086804.

[6] Allen M J, Tung V C, Kaner R B. Honeycomb Carbon: A Review of Graphene[J]. Chemical Reviews, 2010, 110(1): 132-145.

[7] Loh K P, Tong S W, Wu J. Graphene and Graphene-like Molecules: Prospects in Solar Cells[J]. Journal of the American Chemical Society, 2016, 138(4): 1095-1102.

[8] Wang Z, Zhou X-F, Zhang X, Zhu Q, Dong H, Zhao M, Oganov A R. Phagraphene: A Low-Energy Graphene Allotrope Composed of 5–6–7 Carbon Rings with Distorted Dirac Cones[J]. Nano Letters, 2015, 15(9): 6182-6186.

[9] Chen J, Xi J, Wang D, Shuai Z. Carrier Mobility in Graphyne Should Be Even Larger than That in Graphene: A Theoretical Prediction[J]. The Journal of Physical Chemistry Letters, 2013, 4(9): 1443-1448.

[10] Jose D, Datta A. Understanding of the Buckling Distortions in Silicene[J]. The Journal of Physical Chemistry C, 2012, 116(46): 24639-24648.

[11] Cahangirov S, Topsakal M, Aktürk E, Şahin H, Ciraci S. Two- and One-Dimensional Honeycomb Structures of Silicon and Germanium[J]. Physical Review Letters, 2009, 102(23): 236804.

[12] Zhao M, Zhang R. Two-dimensional topological insulators with binary honeycomb lattices: $SiC_3$ siligraphene and its analogs[J]. Physical Review B, 2014, 89(19): 195427.

[13] Lyu J-k, Ji W-x, Zhang S-f, Zhang C-w, Wang P-j. Two-Dimensional Honeycomb $B_2Se$ with Orthogonal Lattice: High Stability and Strong Anisotropic Dirac Cone[J]. The Journal of Physical Chemistry C, 2020, 124(13): 7558-7565.

[14] Zhang Y, Kang J, Zheng F, Gao P-F, Zhang S-L, Wang L-W. Borophosphene: A New Anisotropic Dirac Cone Monolayer with a High Fermi Velocity and a Unique Self-Doping Feature[J]. The Journal of Physical Chemistry Letters, 2019, 10(21): 6656-6663.

[15] Yi W-c, Liu W, Botana J, Zhao L, Liu Z, Liu J-y, Miao M-s. Honeycomb Boron Allotropes with Dirac Cones: A True Analogue to Graphene[J]. The Journal of Physical Chemistry Letters, 2017, 8(12): 2647-2653.

[16] Xu L-C, Wang R-Z, Miao M-S, Wei X-L, Chen Y-P, Yan H, Lau W-M, Liu L-M, Ma Y-M. Two dimensional Dirac carbon allotropes from graphene[J]. Nanoscale, 2014, 6(2): 1113-1118.

[17] Liu C-C, Feng W, Yao Y. Quantum Spin Hall Effect in Silicene and Two-Dimensional Germanium[J]. Physical Review Letters, 2011, 107(7): 076802.

[18] Liu C-C, Jiang H, Yao Y. Low-energy effective Hamiltonian involving spin-orbit coupling in silicene and two-dimensional germanium and tin[J]. Physical Review B, 2011, 84(19): 195430.

[19] Pedersen T G, Fisker C, Jensen R V S. Tight-binding parameterization of α-Sn quasiparticle band structure[J]. Journal of Physics and Chemistry of Solids, 2010, 71(1): 18-23.

[20] Qin X, Wu Y, Liu Y, Chi B, Li X, Wang Y, Zhao X. Origins of Dirac cone formation in $AB_3$ and $A_3B$ (A, B = C, Si, and Ge) binary monolayers[J]. Scientific Reports, 2017, 7(1): 10546.

[21] Chen X, Xu W, Song B, He P. First-principles study of stability, electronic structure and quantum capacitance of B-, N- and O-doped graphynes as supercapacitor electrodes[J]. Journal of Physics: Condensed Matter, 2020, 32(21): 215501.

[22] Bekaroglu E, Topsakal M, Cahangirov S, Ciraci S. First-principles study of defects and adatoms in silicon carbide honeycomb structures[J]. Physical Review B, 2010, 81(7): 075433.

[23] Zhou L-J, Zhang Y-F, Wu L-M. $SiC_2$ Siligraphene and Nanotubes: Novel Donor Materials in Excitonic Solar Cells[J]. Nano Letters, 2013, 13(11): 5431-5436.

[24] Dong H, Zhou L, Frauenheim T, Hou T, Lee S-T, Li Y. $SiC_7$ siligraphene: a novel donor material with extraordinary sunlight absorption[J]. Nanoscale, 2016, 8(13): 6994-6999.

[25] Yu T, Zhao Z, Sun Y, Bergara A, Lin J, Zhang S, Xu H, Zhang L, Yang G, Liu Y. Two-Dimensional $PC_6$ with Direct Band Gap and Anisotropic Carrier Mobility[J]. Journal of the American Chemical Society, 2019, 141(4): 1599-1605.

[26] Kresse G, Hafner J. Ab initio molecular-dynamics simulation of the liquid-metal-amorphous-semiconductor transition in germanium[J]. Physical Review B, 1994, 49(20): 14251-14269.

[27] Kresse G, Furthmüller J. Efficiency of ab-initio total energy calculations for metals and semiconductors using a plane-wave basis set[J]. Computational Materials Science, 1996, 6(1): 15-50.

[28] Kresse G, Furthmüller J. Efficient iterative schemes for ab initio total-energy calculations using a plane-wave basis set[J]. Physical Review B, 1996, 54(16): 11169-11186.

[29] Blöchl P E. Projector augmented-wave method[J]. Physical Review B, 1994, 50(24): 17953-17979.

[30] Perdew J P, Burke K, Ernzerhof M. Generalized Gradient Approximation Made Simple[J]. Physical Review Letters, 1996, 77(18): 3865-3868.

[31] Parlinski K, Li Z Q, Kawazoe Y. First-Principles Determination of the Soft Mode in Cubic $ZrO_2$[J]. Physical Review Letters, 1997, 78(21): 4063-4066.

[32] Heyd J, Scuseria G E, Ernzerhof M. Hybrid functionals based on a screened Coulomb potential[J]. The Journal of Chemical Physics, 2003, 118(18): 8207-8215.

[33] Yi W, Tang G, Chen X, Yang B, Liu X. qvasp: A flexible toolkit for VASP users in materials simulations[J]. Computer Physics Communications, 2020: 107535.

[34] Liu X, Shao X, Yang B, Zhao M. Negative Poisson's ratio and high-mobility transport anisotropy in $SiC_6$ siligraphene[J]. Nanoscale, 2018, 10(4): 2108-2114.

[35] Lu X K, Xin T Y, Zhang Q, Xu Q, Wei T H, Wang Y X. Versatile mechanical properties of novel g-$SiC_x$ monolayers from graphene to silicene: a first-principles study[J]. Nanotechnology, 2018, 29(31): 315701.

[36] Maździarz M. Comment on 'The Computational 2D Materials Database: high-throughput modeling and discovery of atomically thin crystals'[J]. 2D Materials, 2019, 6(4): 048001.

[37] Topsakal M, Cahangirov S, Ciraci S. The response of mechanical and electronic properties of graphane to the elastic strain[J]. Applied Physics Letters, 2010, 96(9): 091912.

[38] Lee C, Wei X, Kysar J W, Hone J. Measurement of the elastic properties and intrinsic strength of monolayer graphene[J]. Science, 2008, 321(5887): 385-388.

[39] Zhang Y, Wu Z-F, Gao P-F, Fang D-Q, Zhang E-H, Zhang S-L. Strain-tunable electronic and optical properties of $BC_3$ monolayer[J]. RSC advances, 2018, 8(3): 1686-1692.

[40] Ding Y, Wang Y. Density Functional Theory Study of the Silicene-like SiX and $XSi_3$ (X = B, C, N, Al, P) Honeycomb Lattices: The Various Buckled Structures and Versatile Electronic Properties[J]. The Journal of Physical Chemistry C, 2013, 117(35): 18266-18278.

[41] Choi S-M, Jhi S-H, Son Y-W. Effects of strain on electronic properties of graphene[J]. Physical Review B, 2010, 81(8): 081407.

[42] Cocco G, Cadelano E, Colombo L. Gap opening in graphene by shear strain[J]. Physical Review B, 2010, 81(24): 241412.

[43] Sun X-H, Li C-P, Wong W-K, Wong N-B, Lee C-S, Lee S-T, Teo B-K. Formation of Silicon Carbide Nanotubes and Nanowires via Reaction of Silicon (from Disproportionation of Silicon Monoxide) with Carbon Nanotubes[J]. Journal of the American Chemical Society, 2002, 124(48): 14464-14471.

**Table of Contents (TOC) Graphic**

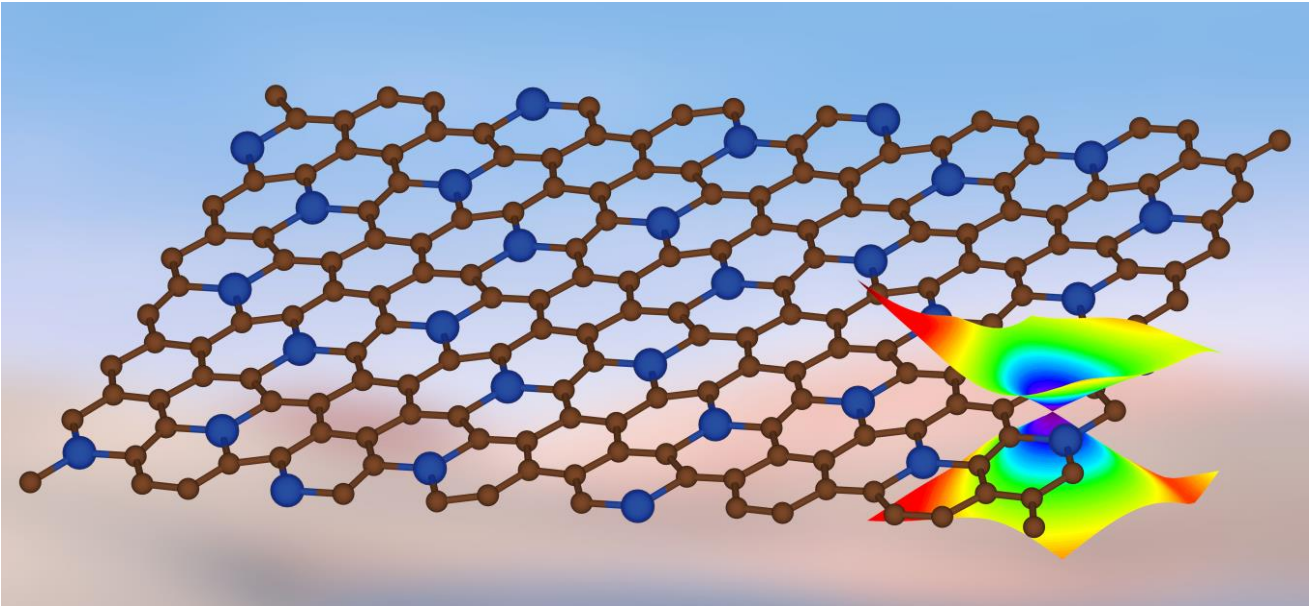

# Supporting Information

# g-SiC$_6$ Monolayer: A New Graphene-like Dirac Cone Material with a High Fermi Velocity

Tao Yang [a,b], Xingang Jiang [a], Wencai Yi [*a], Xiaomin Cheng [*b] and Xiaobing Liu[*a]

c. *Laboratory of High Pressure Physics and Material Science (HPPMS), School of Physics and Physical Engineering, Qufu Normal University, Qufu, Shandong, 273165, China*

d. *Institute of Advanced Materials, School of Electromechanical and Automobile Engineering, Huanggang Normal University, Huanggang, Hubei, 438000, China*

Corresponding Author

* E-mail: yiwc@qfnu.edu.cn (W. Y.)

* E-mail: chengxm@whut.edu.cn (X. C.)

* E-mail: xiaobing.phy@qfnu.edu.cn (X. L.)

Supporting information consists of 8 figures (Figure S1-S8) and 1 table (Table S1).

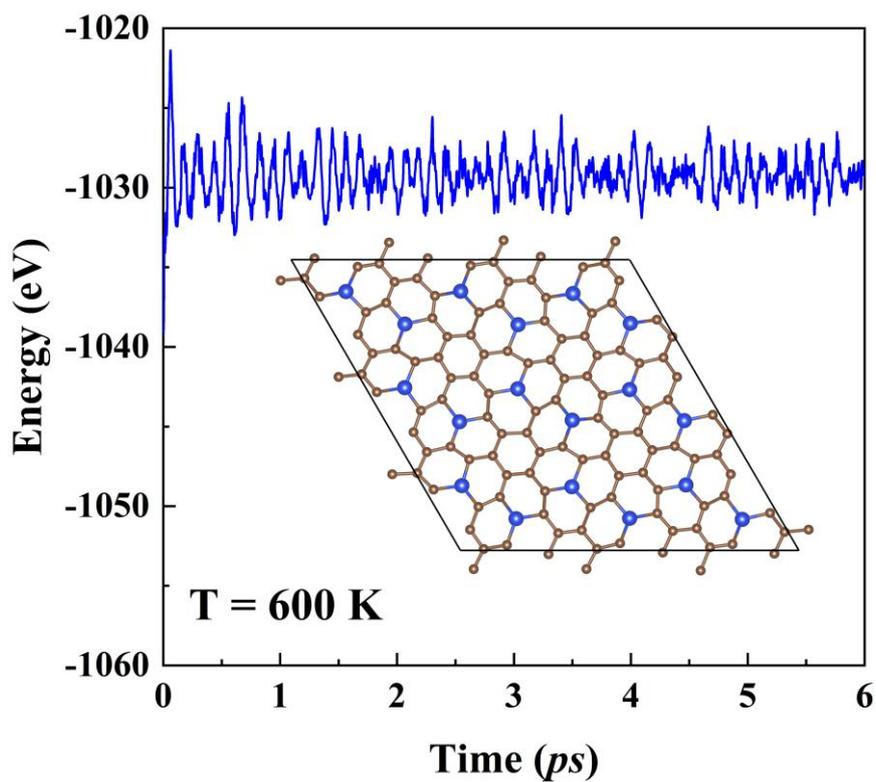

**Figure S1.** Snapshots of g-SiC$_6$ monolayer at the temperatures of 600 K after 6 ps MD simulations.

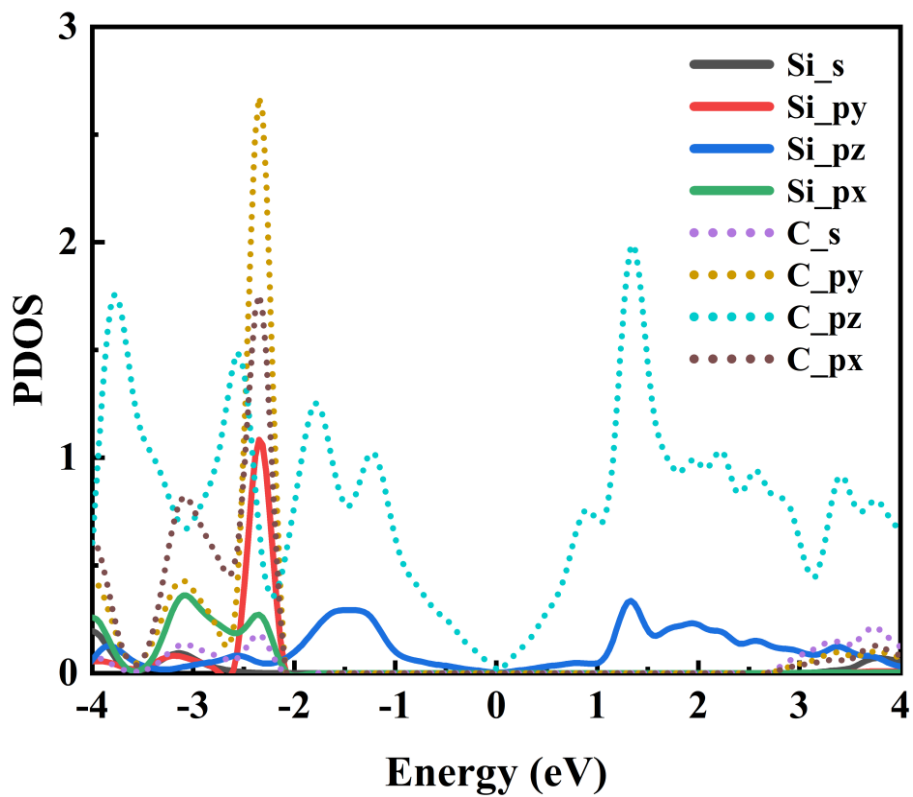

**Figure S2.** Partial density of state (PDOS) of the g-SiC$_6$ monolayer.

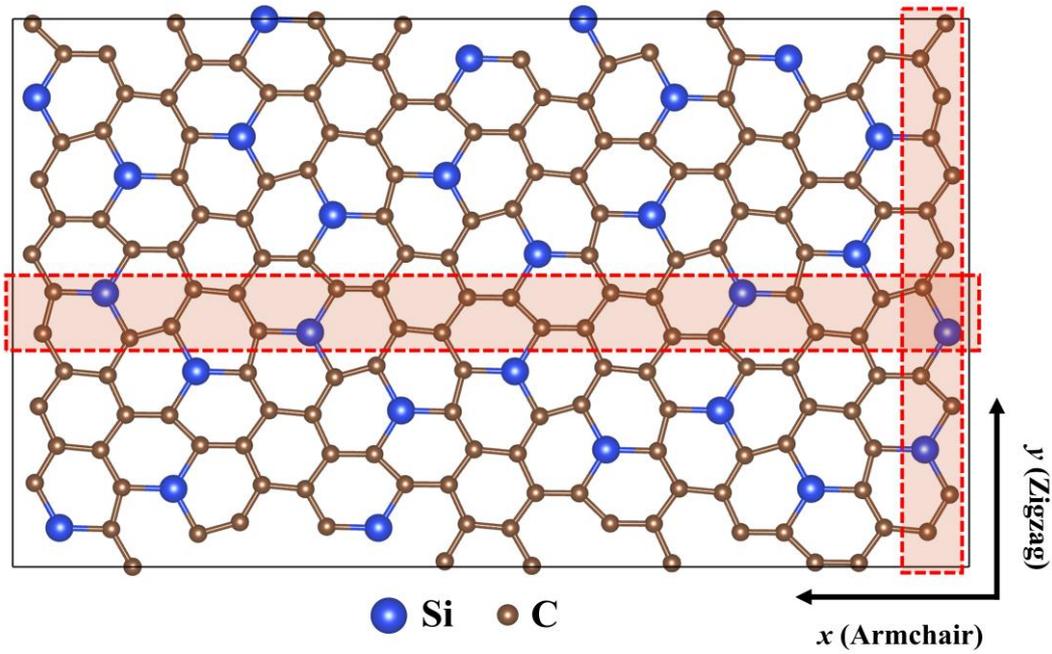

**Figure S3.** A g-SiC$_6$ monolayer in an orthogonal supercell with 196 atoms. It clearly shows that the arrangements of Si and C atoms along zigzag and armchair directions are different.

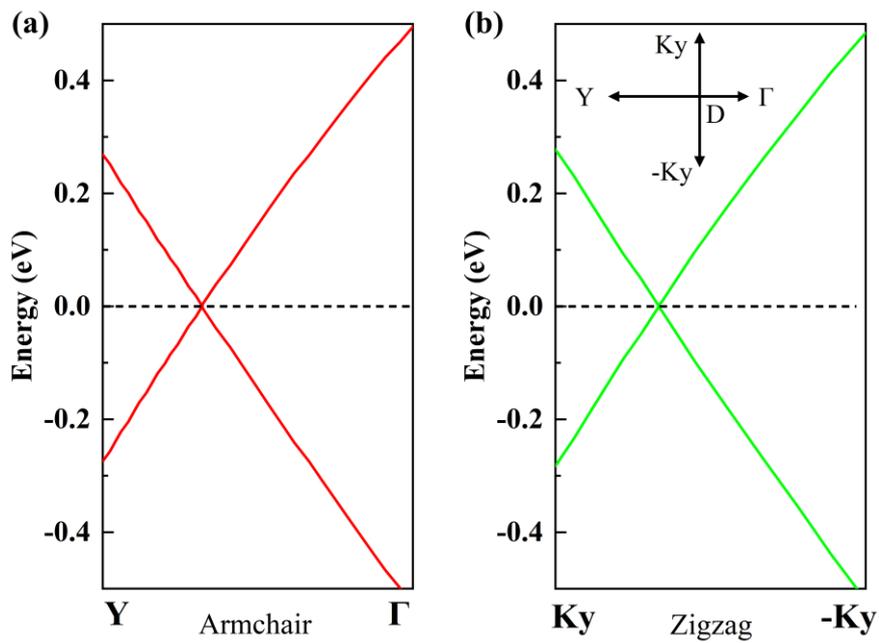

**Figure S4.** Calculated Dirac cones of g-SiC$_6$ monolayer along the (a) armchair ($k_x$) and (b) zigzag ($k_y$) directions.

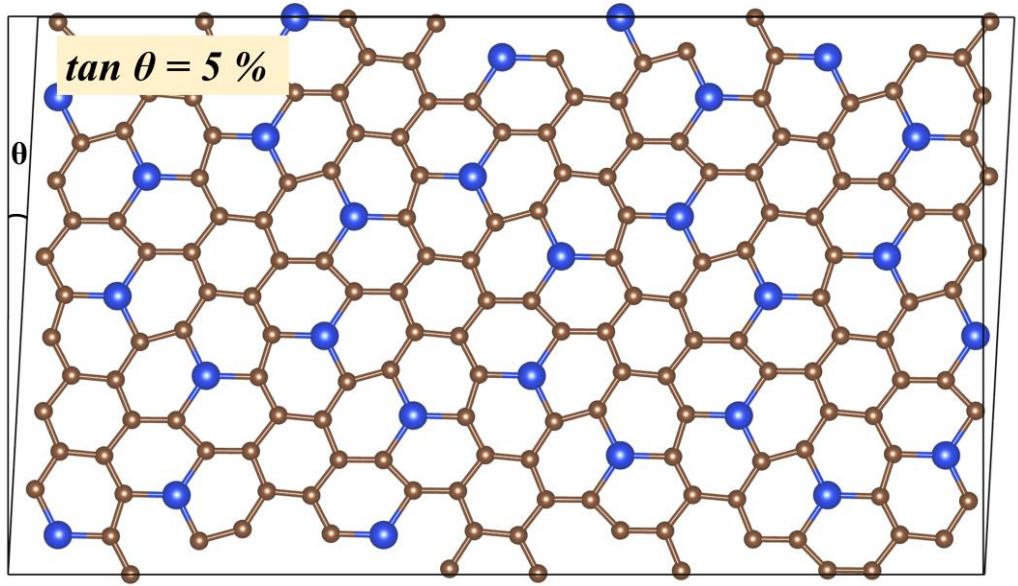

**Figure S5.** The schematic diagram of in-plane shear strain.

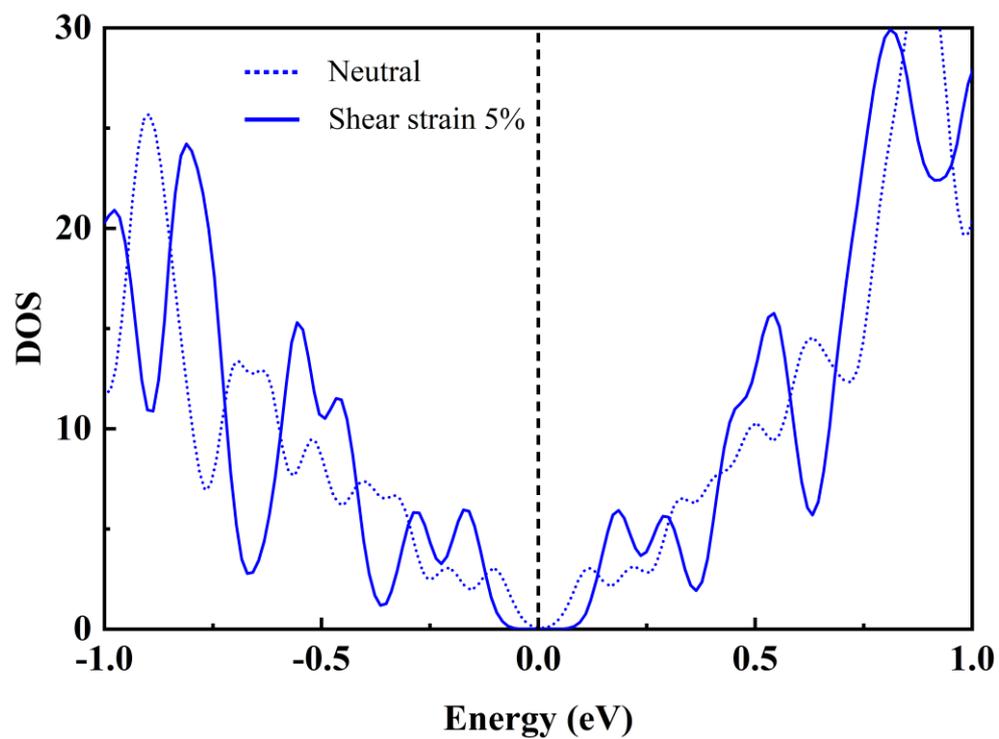

**Figure S6.** The calculated DOS of g-SiC$_6$ monolayer under neutral and shear (5%) strains. It is found that the g-SiC$_6$ monolayer can open a small band gap under a shear strain of 5%.